\documentclass[twocolumn,aps,superscriptaddress,showpacs,nofootinbib,floatfix]{revtex4}

\usepackage{epsfig,bm,feynmf}

\usepackage{graphics}

\usepackage[normalem]{ulem}  
\usepackage[dvips]{color} 

\renewcommand\sout{\bgroup \color{red} \ULdepth=-.5ex \ULset}

\begin{document}



\title{Dilepton production in a schematic causal viscous hydrodynamics}


\author{Taesoo Song}\email{songtsoo@yonsei.ac.kr}
\affiliation{Cyclotron Institute, Texas A$\&$M University, College Station, TX 77843-3366, USA}
\author{Kyong Chol Han}\email{khan@comp.tamu.edu}
\affiliation{Cyclotron Institute and Department of Physics and Astronomy, Texas A$\&$M University, College Station, TX 77843-3366, USA}
\author{Che Ming Ko}\email{ko@comp.tamu.edu}
\affiliation{Cyclotron Institute and Department of Physics and Astronomy, Texas A$\&$M University, College Station, TX 77843-3366, USA}


\begin{abstract}
Assuming that in the hot dense matter produced in relativistic heavy-ion collisions, the energy density, entropy density, and pressure as well as the azimuthal and space-time rapidity components of the shear tensor are uniform in the direction transversal to the reaction plane, we derive a set of schematic equations from the Isreal-Stewart causal viscous hydrodynamics. These equations are then used to describe the evolution dynamics of relativistic heavy-ion collisions by taking the shear viscosity to entropy density ratio of $1/4\pi$ for the initial quark-gluon plasma (QGP) phase and of ten times this value for the later hadron-gas (HG) phase. Using the production rate evaluated with particle distributions that take into account the viscous effect, we study dilepton production in central heavy-ion collisions. Compared with results from the ideal hydrodynamics, we find that although the dilepton invariant mass spectra from the two approaches are similar, the transverse momentum spectra are significantly enhanced at high transverse momenta by the viscous effect. We also study the transverse momentum dependence of dileptons produced from QGP for a fixed transverse mass which is essentially absent in the ideal hydrodynamics, and find that this so-called transverse mass scaling is violated in the viscous hydrodynamics, particularly at high transverse momenta.
\end{abstract}

 \keywords{}

\maketitle


\section{Introduction}

The ideal hydrodynamics without shear viscosity has been quite successful in describing the anisotropic flow of particles in heavy-ion collisions at the Relativistic Heavy Ion Collider (RHIC)~\cite{Huovinen:2001cy,Heinz:2004pj}. Since the viscosity is inversely proportional to the scattering cross section between constituent particles, the applicability of the ideal hydrodynamics at RHIC provides a strong evidence that the quark-gluon plasma formed at RHIC is a strongly coupled one (sQGP). The study based on the Ads/CFT gauge-gravity duality has, however, suggested that the shear viscosity to entropy density in the QGP cannot be smaller than $1/4\pi$. Small viscosities of QGP have also been obtained in studies based on either the quasi-particle model that fits the equation of state from the lattice gauge calculations~\cite{Peshier:2005pp} or the pQCD including both gluon elastic and radiative scatterings that gives a good description of measured elliptic flows at RHIC~\cite{Xu:2007jv}. On the other hand, the viscosity of hadronic matter has been found to be much larger in theoretical studies~\cite{Chen:2006iga,Demir:2008tr,Pal:2010es}, about an order of magnitude larger than the lower bound predicted by the Ads/CFT gauge-gravity duality.

Including a small viscosity in the hydrodynamics has led to an improved description of measured anisotropic flows of hadrons at large transverse momenta~\cite{Romatschke:2007mq}. A nonzero viscosity also affects particle momentum distributions in the hot dense matter produced in heavy-ion collisions, resulting in a deviation from thermal equilibrium during its expansion, and this is particularly so for particles of high transverse momenta. Moreover, the viscosity can change the evolution dynamics of produced hot matter in relativistic heavy-ion collisions. Because of the heat generated by viscosity, cooling of the hot matter becomes slower, leading to a slower decrease of its temperature and thus a larger transversal but a slower longitudinal expansion compared to the case of the ideal hydrodynamics~\cite{Song:2009gc}.

The viscous effect can further affect dilepton production in relativistic heavy-ion collisions, which has been suggested as a possible tool to probe the properties of quark-gluon plasma~\cite{Feinberg:1976ua,Shuryak:1978ij,Ko:1989ek,Xia:1990ps,Asakawa:1993kb,Asakawa:1994nn} as well as those of hot dense hadronic matter~\cite{Xia:1988ym,Li:1995qm,Li:1994cj,Li:1996db,Chung:1998ev,Rapp:1999ej,Cassing:1999es}. As shown in Ref.~\cite{Dusling:2008xj}, the viscosity modifies significantly the transverse momentum spectrum of dileptons produced in relativistic heavy-ion collisions, although not much the invariance mass spectrum. This study was based on the non-causal Navier-Stokes viscous hydrodynamics, which is known to have instabilities in numerical simulations~\cite{Hiscock:1983zz}, and only included the viscous effect on dilepton production due to the leading order correction from modified particle distributions. In the present paper, we extend the study by using the causal Israel-Stewart viscous hydrodynamics and including also the effect of the second-order correction from modified particle distributions on dilepton production. To simplify calculations, we assume that in the hot dense matter produced in relativistic heavy-ion collisions, the energy density, entropy density, and pressure as well as the azimuthal and space-time rapidity components of the shear tensor are uniform in the direction transversal to the reaction plane and derive a set of schematic equations as in Refs.~\cite{Biro:1981es,Ko:1988mf} for the ideal hydrodynamics.

This paper is organized as follows: In Sec.~\ref{hydrodynamics}, general causal viscous hydrodynamic equations are introduced. These equations are simplified in Sec.~\ref{Budapest} by assuming that all thermal quantities as well as the azimuthal and space-time rapidity components of the shear tensor are uniform in the transversal direction. In Sec.~\ref{realistic}, the quasi-particle model of massive quarks and gluons for the quark-gluon plasma and the resonance gas model for the hadronic matter are introduced for describing the equation of state of the hot dense matter produced in relativistic heavy-ion collisions. The schematic viscous hydrodynamic equations are then solved in Sec.~\ref{dynamics} for heavy ion collisions at RHIC energies to show the viscous effect on the evolution dynamics and the distribution of particle transverse momentum spectra. In Sec.~\ref{dilepton}, dilepton production is studied in the viscous hydrodynamics to find out the effect of viscosity on the dilepton invariant mass and transverse momentum spectra. Finally, discussions and summary are given in Sec.~\ref{summary}. Details on the derivation of the dilepton production rate using the modified particle distributions from the viscous hydrodynamics is given in the Appendix.

\section{the viscous hydrodynamics}\label{hydrodynamics}

In hydrodynamic description of relativistic heavy-ion collisions, the hot dense matter is characterized by its net charge currents and energy-momentum tensor. Since particles at midrapidities are largely produced ones, their net charge currents are essentially zero and can be safely neglected. In the Landau and Lifshitz frame, which assumes that the four-vector velocity $u^\mu=\gamma(1,{\vec v})$ is parallel to the energy flow and the heat conductivity is zero~\cite{Rishke:1998}, the energy-momentum tensor can be written as~\cite{Heinz:2005bw}
\begin{eqnarray}
T^{\mu\nu}=(e+p)u^\mu u^\nu-pg^{\mu\nu}+\pi^{\mu\nu},
\end{eqnarray}
where $e$ and $p$ are the energy density and pressure, respectively, and $\pi^{\mu\nu}$ is the traceless symmetric shear tensor. At  midrapidity particles follow essentially the boost-invariant expansion along the longitudinal direction~\cite{Bjorken:1982qr}, i.e., the longitudinal flow velocity is equal to $z/t$, if it starts at $z=t=0$. Furthermore, the transverse flow velocity is independent of the azimuthal angle $\phi$ in central heavy-ion collisions. In the $(\tau, r, \phi, \eta)$ coordinate system defined by
\begin{eqnarray}
\tau&=&\sqrt{t^2-z^2}, ~~\eta=\frac{1}{2}\ln \frac{t+z}{t-z},\nonumber\\
r&=&\sqrt{x^2+y^2}, ~~\phi=\tan^{-1}(y/x),
\end{eqnarray}
only $T^{\tau\tau}$, $T^{rr}$, $T^{\tau r}$, $T^{\eta\eta}$ and $T^{\phi\phi}$ components of the energy-momentum tensor, and $\pi^{\tau\tau}$, $\pi^{rr}$, $\pi^{\tau r}$, $\pi^{\eta\eta}$ and $\pi^{\phi\phi}$ of the shear tensor are non-zero in central heavy-ion collisions. For the energy-momentum tensor, they are given by
\begin{eqnarray}
T^{\tau\tau}&=&(e+P_r)u_\tau^2 -P_r\nonumber\\
T^{\tau r}&=&(e+P_r)u_\tau u_r\nonumber\\
T^{r r}&=&(e+P_r)u_r^2+P_r
\label{tensor6}
\end{eqnarray}
where $P_r\equiv p-\tau^2\pi^{\eta\eta}-r^2\pi^{\phi\phi}$ is the effective radial pressure. The azimuthal and space-time components of the shear tensor $r^2\pi^{\phi\phi}$ and $\tau^2\pi^{\eta\eta}$ are the only independent ones as the others can be related to them according to

\begin{eqnarray}
\pi^{\tau r}&=&v_r\pi^{rr}\nonumber\\
\pi^{\tau\tau}&=&v_r\pi^{\tau r}=v_r^2\pi^{rr}\nonumber\\
\pi^{rr}&=&-\gamma_r^2(r^2\pi^{\phi\phi}+\tau^2\pi^{\eta\eta}),
\label{shear}
\end{eqnarray}
where the first two equations are derived from $u_\mu\pi^{\mu\nu}=0$ and the last one from the traceless property $\pi_\mu^\mu=0$. The shear tensor components $\pi^{\phi\phi}$ and $\pi^{\eta\eta}$ are boost-invariant in the radial direction and satisfy following simplified Israel-Stewart equations:~\cite{Heinz:2005bw}
\begin{eqnarray}
(\partial_\tau +v_r \partial_r)\pi^{\eta \eta}=-\frac{1}{\gamma_r \tau_\pi}\bigg[\pi^{\eta \eta}-\frac{2\eta_s}{\tau^2}\bigg(\frac{\theta}{3}-\frac{\gamma_r}{\tau}\bigg)\bigg],\label{shear1a}\\
(\partial_\tau +v_r \partial_r)\pi^{\phi \phi}=-\frac{1}{\gamma_r \tau_\pi}\bigg[\pi^{\phi \phi}-\frac{2\eta_s}{r^2}\bigg(\frac{\theta}{3}-\frac{\gamma_r v_r}{r}\bigg)\bigg],
\label{shear1b}
\end{eqnarray}
where
\begin{eqnarray}
\theta=\partial\cdot u=\frac{1}{\tau}\partial_\tau (\tau \gamma_r)+\frac{1}{r}\partial_r(rv_r \gamma_r)\nonumber
\end{eqnarray}
with $\eta_s$ and $\tau_\pi$ being the shear viscosity and the relaxation time for the particle distributions, respectively.

From the energy-momentum conservation conditions $\partial_\mu T^{\mu\nu}=0$, we then obtain following viscous hydrodynamic equations for the produced fire-cylinder:
\begin{eqnarray}
\frac{1}{\tau}\partial_\tau(\tau T^{\tau \tau})+\frac{1}{r}\partial_r(r T^{r \tau})&=&-\frac{1}{\tau}(p+\tau^2\pi^{\eta\eta}),
\label{energy6}\\
\frac{1}{\tau}\partial_\tau(\tau T^{\tau r})+\frac{1}{r}\partial_r(r T^{r r})&=&\frac{1}{r}(p+r^2\pi^{\phi\phi}).
\label{momentum6}
\end{eqnarray}
Furthermore, the condition $u_\mu (T_{;\nu}^{\nu \mu})=0$, where the flow velocity $(u_\tau, u_r, u_\phi, u_\eta)=(\gamma/\cosh\eta,\gamma v_r,0,0)$ reduces to $(\gamma_r,\gamma_r v_r,0,0)$ with $\gamma_r=1/\sqrt{1-v_r^2}$ in midrapidities, leads to
\begin{eqnarray}
\frac{1}{\tau}\partial_\tau (\tau s \gamma_r)+\frac{1}{r}\partial_r (rs\gamma_r v_r)=-\frac{1}{T}\bigg[\frac{u_\tau}{\tau}\tau^2\pi^{\eta\eta}
\nonumber\\
+\frac{u_r}{r}r^2\pi^{\phi\phi}-(\partial_\tau u_\tau+\partial_r u_r)(r^2\pi^{\phi\phi}+\tau^2\pi^{\eta \eta})\bigg],
\label{entropy6}
\end{eqnarray}
where $s=(e+p)/T$ is the local entropy density in the hot dense matter.
Eq. (\ref{entropy6}) shows that a nonzero shear tensor affects the entropy density of the matter.

\section{A schematic viscous hydrodynamics}\label{Budapest}

If all thermal quantities like energy density, temperature, entropy density, and pressure as well as the azimuthal and space-time components of the shear tensor are uniform along the transverse direction in the hot dense matter produced in heavy-ion collisions, we can then simplify the causal viscous hydrodynamic equations by integrating them over the transverse area~\cite{Biro:1981es,Ko:1988mf}. Specifically, we integrate Eqs. (\ref{energy6}) and (\ref{entropy6}) as well as Eqs. (\ref{shear1a}) and (\ref{shear1b}) multiplied by $\gamma_r \tau^2$ and $\gamma_r r^2$, respectively, over the transverse area. In terms of $\pi^\eta_\eta=\tau^2\pi^{\eta\eta}$ and $\pi^\eta_\eta=\tau^2\pi^{\eta\eta}$, this leads to
\begin{eqnarray}
&&\partial_\tau (A\tau \langle T^{\tau \tau}\rangle)=-(p+\pi^\eta_\eta)A,\label{energy7}\\
\nonumber\\
&&\frac{T}{\tau}\partial_\tau (A\tau s \langle \gamma_r\rangle)=-A\bigg\langle\frac{\gamma_r v_r}{r}\bigg\rangle \pi^\phi_\phi-\frac{A\langle \gamma_r\rangle}{\tau}\pi^\eta_\eta\nonumber\\
&&~~~~~~~+\bigg\{\partial_\tau(A\langle \gamma_r\rangle)-\frac{\gamma_R \dot{R}}{R}A\bigg\}(\pi^\phi_\phi+\pi^\eta_\eta),\label{entropy7}\\
\nonumber\\
&&\partial_\tau (A\langle \gamma_r\rangle \pi^\eta_\eta) -\bigg\{\partial_\tau(A\langle\gamma_r\rangle)+2\frac{A\langle\gamma_r\rangle}{\tau} \bigg\}\pi^\eta_\eta\nonumber\\
&&~~~~=-\frac{A}{\tau_\pi}\bigg[\pi^\eta_\eta-2\eta_s\bigg\{\frac{\langle\theta\rangle}{3}-\frac{\langle\gamma_r\rangle}{\tau}\bigg\}\bigg],\label{entropy7}\\
\nonumber\\
&&\partial_\tau(A\langle\gamma_r\rangle~ \pi^\phi_\phi)-\bigg\{\partial_\tau(A\langle\gamma_r\rangle)+2A\bigg\langle\frac{\gamma_r v_r}{r}\bigg\rangle\bigg\}\pi^\phi_\phi\nonumber\\
&&~~~~=-\frac{A}{\tau_\pi}\bigg[ \pi^\phi_\phi-2\eta_s \bigg\{\frac{\langle\theta\rangle}{3}-\bigg\langle\frac{\gamma_r v_r}{r}\bigg\rangle\bigg\}\bigg]\label{shear7b},
\end{eqnarray}
where $A=\pi R^2$ with $R$ being the transverse radius of the uniform matter and $\langle\cdots\rangle$ denotes average over the transverse area. Assuming that
the radial flow velocity is a linear function of the radial distance from the center, i.e.,
$\gamma_r v_r=\gamma_R \dot{R}(r/R)$, where $\dot{R}=\partial R/\partial \tau$ and $\gamma_R=1/\sqrt{1-\dot{R}^2}$, we then have
\begin{eqnarray}
&&\langle\gamma_r^2\rangle=1+\frac{\gamma_R^2 \dot{R}^2}{2}\nonumber\\
&&\langle\gamma_r^2 v_r^2\rangle=\frac{\gamma_R^2 \dot{R}^2}{2}\nonumber\\
&&\langle\gamma_r\rangle=\frac{2}{3\gamma_R^2 \dot{R}^2}\left(\gamma_R^3-1\right)\nonumber\\
&&\bigg\langle\frac{\gamma_r v_r}{r}\bigg\rangle=\frac{\gamma_R \dot{R}}{R}.
\label{gamma}
\end{eqnarray}
Since the energy density $e$ and pressure $p$ are related by the equation of state of the matter through its temperature $T$, Eqs.(\ref{energy7})-(\ref{shear7b}) are thus four simultaneous equations for $T$, $\dot{R}$, $\pi^\phi_\phi$ and $\pi^\eta_\eta$.

\section{the equation of state}\label{realistic}

For the equation of state of QGP, we use the quasi-particle model of Ref.~\cite{Levai:1997yx}, which assumes that the QGP is composed of noninteractig massive quarks and gluons. In terms of the temperature $T$ of QGP, their masses are given by
\begin{eqnarray}
m_g^2&=&\bigg(\frac{N_c}{3}+\frac{N_f}{6}\bigg)\frac{g^2(T)T^2}{2}, \nonumber\\
m_q^2&=&\frac{g^2(T)T^2}{3},
\end{eqnarray}
where the strong coupling constant $g(T)$ is given by
\begin{eqnarray}
g^2(T)&=&\frac{48\pi^2}{(11N_c-2N_f)\ln F^2(T,T_c,\Lambda)}, \nonumber\\
F(T,T_c,\Lambda)&=&\frac{18}{18.4e^{(T/T_c)^2/2}+1}\frac{T}{T_c}\frac{T_c}{\Lambda},\nonumber
\end{eqnarray}
with $T_c=170$ MeV, $T_c/\Lambda=1.05$, $N_c=3$ and $N_f=3$. The
pressure, energy density, and entropy density of QGP are then given, respectively, by
\begin{eqnarray}
p(T)&=& \sum_i \frac{g_i}{6\pi^2}\int^\infty_0 dk f_i(T) \frac{k^4}{E_i}-B(T)\nonumber\\
&\equiv& p_0(T)-B(T)\nonumber\\
e(T)&=& \sum_i \frac{g_i}{2\pi^2}\int^\infty_0 dk k^2 f_i(T) E_i +B(T)\nonumber\\
s(T)&=& \sum_i \frac{g_i}{2\pi^2T}\int^\infty_0 dk f_i(T) \frac{\frac{4}{3}k^2+m_i^2(T)}{E_i},
\label{eps}
\end{eqnarray}
with $m_i(T)$ and $g_i$ being, respectively, the thermal mass and degeneracy factor of parton species $i$.
The parton distribution function is denoted by
\begin{equation}
f_i(T)=\frac{1}{e^{E_i/T}\pm 1}
\end{equation}
with the plus and minus signs in the denominator for quarks and gluons, respectively,
and $E_i=\sqrt{m_i^2+k^2}$. For the bag pressure
$B(T)$, it is determined from the relation $s=\partial p/\partial T$ such that
\begin{eqnarray}
B(T)=B_0 +\sum_i \int_{T_c}^T dT \frac{\partial p_0}{\partial m_i^2(T)}\frac{\partial m_i^2(T)}{\partial T},
\end{eqnarray}
where $B_0$ is the bag pressure at $T_c$ and is taken to be 0.095 times the energy density at this temperature in order to keep the pressure continuous at $T_c$. A similar value of $B_0$ has been used in Ref.~\cite{Song:2010er} for the case of nozero but small baryon chemical potential.

For the HG phase, we use the resonance gas model that includes both stable hadrons and their resonances up to 1.5 GeV for mesons and 2.0 GeV for baryons. Its pressure, energy density and entropy density can be similarly evaluated as those in Eq. (\ref{eps}) for the QGP, except that the bag constant is not present and the hadron masses are taken to be their values in free space.

Because of the larger entropy density in QGP than in HG at $T_c$, a mixed phase of constant temperature $T_c$ is introduced during the transition between these two phases of matter. In terms of the fraction $f$ of HG in the mixed phase, the entropy of the mixed phase is
\begin{eqnarray}
s=fs^H+(1-f)s^Q,
\end{eqnarray}
where $s^H$ and $s^Q$ are, respectively, the entropy density of HG and QGP at $T_c$.
Similar relations hold for the energy density and pressure.

We refer the reader to Ref.~\cite{Song:2010ix} for details of the above equations of state for QGP and HG.

For the shear viscosity, we take its ratio with respect to the entropy density to be $1/4\pi$ for QGP as given by the Ads/CFT gauge-gravity duality~\cite{Kovtun:2004de} and ten times this value for HG as determined from the hadronic transport model~\cite{Demir:2008tr}. For the relaxation time $\tau_\pi$ in Eqs. (\ref{shear1a}) and (\ref{shear1b}), we use the assumption $\eta/\tau_\pi=sT/3$~\cite{Song:2009gc} for both QGP and HG.

\section{Heavy-ion collision dynamics in the schematic viscous hydrodynamics}\label{dynamics}

To apply the schematic viscous hydrodynamic equations, Eqs. (\ref{energy7})-(\ref{shear7b}), to heavy-ion collisions, we
divide the time into infinitesimal intervals. These equations then become
\begin{widetext}
\begin{eqnarray}
&&\langle\gamma_r^2\rangle_{n+1}e_{n+1}+\langle\gamma_r^2v_r^2\rangle_{n+1}\bigg[p_{n+1}-(\pi^\eta_\eta)_{n+1}-(\pi^\phi_\phi)_{n+1}\bigg]\nonumber\\
&&~~=\frac{R_n^2\tau_n}{R_{n+1}^2\tau_{n+1}}\bigg[\langle\gamma_r^2\rangle_{n}e_{n} +\bigg\{\langle\gamma_r^2v_r^2\rangle_{n}-\frac{\Delta\tau}{\tau_n}\bigg\}p_{n}
-\bigg\{\langle\gamma_r^2v_r^2\rangle_{n}+\frac{\Delta\tau}{\tau_n}\bigg\}(\pi^\eta_\eta)_n-\langle\gamma_r^2v_r^2\rangle_{n}(\pi^\phi_\phi)_{n}\bigg],\label{energy8}\\
\nonumber\\
&&\langle\gamma_r\rangle_{n+1}\bigg[\frac{\tau_{n+1}}{\tau_n}T_n s_{n+1}-(\pi^\phi_\phi)_n-(\pi^\eta_\eta)_n\bigg]\nonumber\\
&&~~=\frac{R_n^2}{R_{n+1}^2}\langle\gamma_r\rangle_n\bigg[T_n s_n -\bigg\{1+\frac{2(\gamma_R\dot{R})_n \Delta \tau}{\langle\gamma_r\rangle_n R_n}\bigg\}(\pi^\phi_\phi)_n-\bigg\{1+\frac{(\gamma_R\dot{R})_n \Delta \tau}{\langle\gamma_r\rangle_n R_n}+\frac{\Delta \tau}{\tau_n}\bigg\}(\pi^\eta_\eta)_n\bigg],\label{entropy8}\\
\nonumber\\
&&\langle\gamma_r\rangle_{n+1} \bigg[(\pi^\eta_\eta)_{n+1}-(\pi^\eta_\eta)_n-\frac{2}{3}\bigg(\frac{\eta_s}{\tau_\pi}\bigg)_n \bigg]\nonumber\\
&&~~=\frac{R_n^2\Delta\tau}{R_{n+1}^2}\bigg[\bigg\{\frac{2\langle\gamma_r\rangle_n}{\tau_n}-\frac{1}{(\tau_\pi)_n} \bigg\}(\pi^\eta_\eta)_n -\frac{2}{3}\bigg(\frac{\eta_s}{\tau_\pi}\bigg)_n \bigg(\frac{1}{\Delta\tau}+\frac{2}{\tau_n}\bigg)\langle\gamma_r\rangle_n\bigg],\label{shear8a}\\
\nonumber\\
&&\langle\gamma_r\rangle_{n+1} \bigg[(\pi^\phi_\phi)_{n+1}-(\pi^\phi_\phi)_n-\frac{2}{3}\bigg(\frac{\eta_s}{\tau_\pi}\bigg)_n \bigg]\nonumber\\
&&~~=\frac{R_n^2\Delta\tau}{R_{n+1}^2}\bigg[\bigg\{2\bigg\langle\frac{\gamma_r v_r}{r}\bigg\rangle_n -\frac{1}{(\tau_\pi)_n} \bigg\}(\pi^\phi_\phi)_n -\frac{2}{3}\bigg(\frac{\eta_s}{\tau_\pi}\bigg)_n \bigg\{\bigg(\frac{1}{\Delta\tau}-\frac{1}{\tau_n}\bigg)\langle\gamma_r\rangle_n +3\bigg\langle\frac{\gamma_r v_r}{r}\bigg\rangle_n \bigg\}\bigg]\label{shear8b},
\end{eqnarray}
\end{widetext}
where $\Delta\tau=\tau_{n+1}-\tau_n$ with the subscript denoting the time step and $(\partial O/\partial \tau)_n=(O_{n+1}-O_n)/\Delta\tau$.

\begin{figure}[h]
\centerline{
\includegraphics[width=8.5 cm]{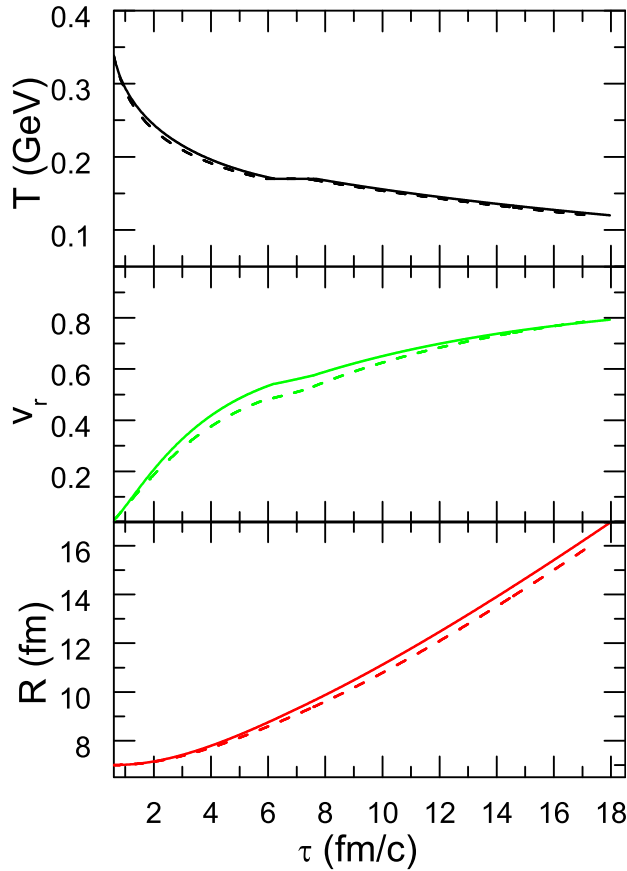}}
\caption{Time evolutions of temperature $T$ (top panel), radial flow velocity $V_r$ (middle panel), and transverse radius $R$ (bottom panel) of the fire-cyclinder in viscous (solid lines) and ideal (dashed lines) hydrodynamics.}
\label{real-viscous1}
\end{figure}

Eqs. (\ref{energy8})-(\ref{shear8b}) are solved by using following relations between the $(n+1)$th and $n$th time steps in the energy density, pressure and entropy:
\begin{eqnarray}
e_{n+1}&=&e_n+\frac{\partial e}{\partial T}\bigg|_n (T_{n+1}-T_n),\nonumber\\
p_{n+1}&=&p_n+\frac{\partial p}{\partial T}\bigg|_n (T_{n+1}-T_n),\nonumber\\
s_{n+1}&=&s_n+\frac{\partial s}{\partial T}\bigg|_n (T_{n+1}-T_n),
\label{eps2}
\end{eqnarray}
where
\begin{eqnarray}
\frac{\partial s}{\partial T}&=&-\frac{s}{T}+\sum_i \frac{g_i}{2\pi^2 T}\int dk\bigg[\frac{\partial f_i}{\partial T}\frac{(4/3)k^2+m_i^2}{E_i}\nonumber\\
&&~~~~~~~~~~+f_i\frac{(4/3)k^2+\partial m_i^2/\partial T}{E_i}\bigg]\nonumber\\
\frac{\partial e}{\partial T}&=&\sum_i\frac{g_i}{2\pi^2}\int dk k^2\bigg[\frac{\partial f_i}{\partial T}E_i+\frac{f_i}{2E_i}\frac{\partial m_i^2}{\partial T}\bigg]\nonumber\\
&&~~~~~~~~~~+\frac{\partial p_0}{\partial m_i^2}\frac{\partial m_i^2}{\partial T}\nonumber\\
\frac{\partial p}{\partial T}&=&\sum_i\frac{g_i}{6\pi^2}\int dk \frac{k^4}{E_i}\bigg[\frac{\partial f_i}{\partial T}-\frac{f_i}{2E_i^2}\frac{\partial m_i^2}{\partial T}\bigg]\nonumber\\
&&~~~~~~~~~~-\frac{\partial p_0}{\partial m_i^2}\frac{\partial m_i^2}{\partial T}.\nonumber
\end{eqnarray}

In the following, we show results obtained from the schematic viscous hydrodynamics with the initial conditions of $\tau_0=$ 0.6 fm/$c$ for the thermalization time, $T_0=$ 338 MeV for the initial temperature, and $V_0=0.01~c$ for the initial radial flow velocity~\cite{Song:2010ix}, which are appropriate for heavy-ion collisions at the top RHIC energy. In the top panel of Fig. \ref{real-viscous1}, the time evolution of the temperature $T$ of the fire-cylinder is given by the solid line. It is seen that the QGP, mixed, and HG phases last for about 5.6 fm/$c$, 1.5 fm/$c$, and 10.2 fm/$c$, respectively, which are similar to those in the ideal hydrodynamics with zero viscosities in both QGP and HG as shown by the dashed line in the top panel of Fig. \ref{real-viscous1}. For the time evolutions of the radial flow velocity $V_R=\dot{R}$ and transverse radius $R$ of the fire-cylinder in viscous hydrodynamics, they are shown by solid lines in the middle and bottom panels of Fig.~\ref{real-viscous1}, respectively, and both are slightly above those in ideal hydrodynamics (dashed lines).

\begin{figure}[h]
\centerline{
\includegraphics[width=8.5 cm]{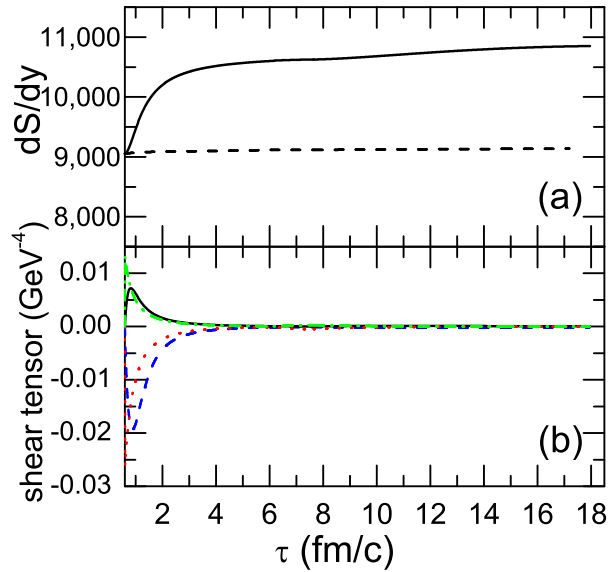}}
\caption{(a) Entropy per unit rapidity in viscous (solid line) and ideal (dashed line) hydrodynamics. (b) Shear tensor components $\pi^\phi_\phi$ (solid line) and $\pi^\eta_\eta$ (dashed line) in the viscous hydrodynamics.
Dash-dotted and dotted lines are $\pi^\phi_\phi$ and $\pi^\eta_\eta$ in the Navier-Stokes limit.}
\label{real-viscous2}
\end{figure}

Fig. \ref{real-viscous2}(a) shows the change of entropy per unit rapidity in time. Because of non-zero viscosity, the entropy per unit rapidity (solid line) increase with time in the viscous hydrodynamics, reaching a value at thermal freeze out which is significantly larger than that in the ideal hydrodynamics (dashed line). Fig. \ref{real-viscous2}(b) shows the time evolution of the shear tensor components $\pi^\phi_\phi$ (solid line) and $\pi^\eta_\eta$ (dashed line). The two are seen to satisfy approximately the relation $\pi^\phi_\phi\approx -\pi^\eta_\eta/2$. Using this result in $\pi^\phi_\phi+\pi^\eta_\eta=-\pi^{rr}$, which is due to traceless of the shear tensor, leads to $\pi^\phi_\phi\approx\pi^{rr}$. Since the two are equal in the absence of radial flow~\cite{Fries:2008}, our results thus indicate that the assumption of uniform $\pi^\phi_\phi$ and $\pi^\eta_\eta$ in central heavy-ion collisions is reasonable.

For comparisons, we also show in Fig. \ref{real-viscous2}(b) the shear tensor components $\pi^\phi_\phi$ (dash-dotted line) and $\pi^\eta_\eta$ (dotted line) obtained in the Navier-Stokes limit, i.e.,
\begin{eqnarray}
\pi^\eta_\eta&=&2\eta_s\bigg[\frac{1}{3\tau A}\partial_\tau (\tau A\langle\gamma_r\rangle)-\frac{\langle\gamma_r\rangle}{\tau}\bigg],\\
\pi^\phi_\phi&=&2\eta_s\bigg[\frac{1}{3\tau A}\partial_\tau (\tau A\langle\gamma_r\rangle)-\bigg\langle\frac{\gamma_r v_r}{r}\bigg\rangle\bigg].
\end{eqnarray}
It is seen that they only differ appreciably from those in the Israel-Stewart causal viscous hydrodynamics at early times.

\begin{figure}[h]
\centerline{
\includegraphics[width=8.5 cm]{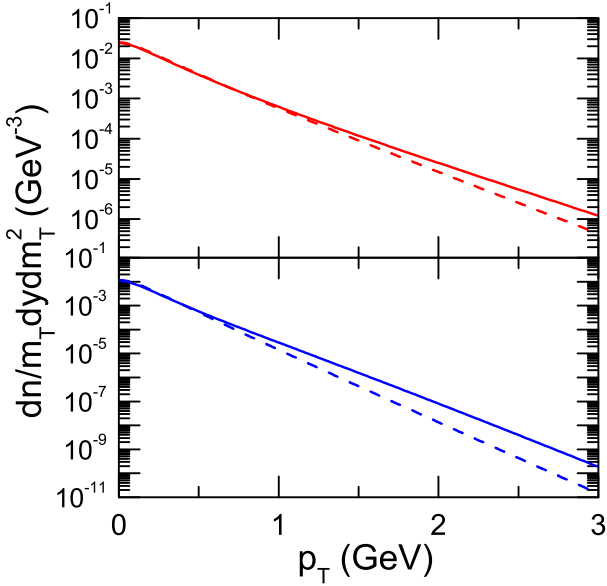}}
\caption{Quark (top panel) and pion (lower panel) transverse momentum spectra from the viscous (solid lines) and ideal (dashed lines) hydrodynamics at T=300 MeV and 150 MeV, respectively.}
\label{spectra}
\end{figure}

Although the evolution dynamics in the viscous hydrodynamics does not differ very much from that in the ideal hydrodynamics, the transverse momentum spectra of particles can differ significantly. In Fig.~\ref{spectra}, we show the quark (top panel) and pion (lower panel) transverse momentum spectra from the viscous (solid lines) and ideal (dashed lines) hydrodynamics at temperatures of 300 and 150 MeV, respectively. It is seen that the transverse momentum spectra from the viscous hydrodynamics are enhanced at high transverse momenta compared to those from the ideal hydrodynamics. The inverse slope parameters of the quark and pion transverse momentum spectra are 330 and 160 MeV, respectively, which are significantly larger than those in the ideal hydrodynamics. The total particle density is, however, not much affected by the viscosity. To understand these results, we note that the particle transverse momentum spectrum in the viscous hydrodynamics is given by~\cite{Baier:2006um}
\begin{eqnarray}
f(k)=f_0(k)+\delta f(k)=f_0(k)\bigg[1+\frac{p_\mu p_\nu \pi^{\mu \nu}}{2T(e+p)}\bigg],
\label{vdis}
\end{eqnarray}
where $f_0(k)$ is the equilibrium thermal distribution of particles in the ideal hydrodynamics and $\delta f(k)$ is the viscous correction. In the ($\tau, r, \phi, \eta$) coordinate system, the factor $p_\mu p_\nu \pi^{\mu \nu}$ can be written as~\cite{Baier:2006gy},
\begin{eqnarray}
p_\mu p_\nu \pi^{\mu \nu}=m_T^2\cosh^2(y-\eta)\pi^{\tau\tau} +p_T^2\cos^2(\phi_p-\phi)\pi^{rr}\nonumber\\ +p_T^2\sin^2(\phi_p-\phi)\pi^\phi_\phi +m_T^2\sinh^2(y-\eta)\pi^\eta_\eta\nonumber\\
-2m_Tp_T\cosh(y-\eta)\cos(\phi_p-\phi)\pi^{\tau r},~~~~
\label{distri1}
\end{eqnarray}
where $m_T=\sqrt{m^2+p_T^2}$ with $p_T$ being the transverse momentum, and $y$ and $\phi_p$ are the energy-momentum rapidity and the azimuthal angle of the momentum, respectively. In the absence of radial flow,
Eq. (\ref{distri1}) reduces to
\begin{eqnarray}
p_\mu p_\nu \pi^{\mu \nu}=p_T^2\pi^{rr}+p_L^2 \pi^\eta_\eta.
\label{shear2}
\end{eqnarray}
Since $\pi^{rr}$ is positive, the particle transverse momentum distribution is thus enhanced at high $p_T$. On the other hand, the particle number density is not much affected by the viscosity as all components of the shear tensor have similar magnitude due to its traceless property.

\section{Dilepton production in the viscous hydrodynamics}\label{dilepton}

The dilepton production rate from the scattering of two particles in hot dense matter is given by
\begin{eqnarray}
\frac{dN}{d^4x}&=&\int \frac{d^3{\bf k}_1}{(2\pi)^3}\frac{d^3{\bf k}_2}{(2\pi)^3}f(k_1)f(k_2)v_{\rm rel}\sigma,
\label{rate1}
\end{eqnarray}
where ${\bf k}_1$ and ${\bf k}_2$ are momenta of the two particles; $v_{\rm rel}$ is their relative velocity; and $\sigma$ is the cross section for dilepton production from their scattering.
For QGP, we consider the dominant quark-antiquark annihilation process for dilepton production, and its cross section is
\begin{eqnarray}
\sigma(M^2)=\frac{4\pi \alpha^2}{3N_c}\sum_{i=u,d,s}\frac{e_i^2}{e^2}\frac{1+2m_q^2/M^2}{M^2\sqrt{1-4m_q^2/M^2}},
\end{eqnarray}
where $m_q$ is the quark mass, $e_i$ is the charge of quark species $i$, and $M^2=(k_1+k_2)^2$ is the squared invariant mass. For HG, the dominant pion-pion annihilation is considered, and the cross section is
\begin{eqnarray}
\sigma(M^2)=\frac{4\pi\alpha^2}{3}\frac{|F(M^2)|^2}{M^2}\sqrt{1-\frac{4m_{\pi}^2}{M^2}},
\end{eqnarray}
with the electromagnetic form factor of pion
\begin{eqnarray}
|F_\pi (M^2)|^2=\sum_{i=\rho,\rho ',\rho ''} \frac{N_i m^4_i}{(m_i^2-M^2)^2+m_i^2\Gamma_i^2},
\end{eqnarray}
where $\rho$, $\rho'$, and $\rho''$ denote $\rho(770)$, $\rho'(1450)$, and $\rho''(1700)$, respectively with their respective width of $\Gamma_\rho=$153 MeV, $\Gamma_{\rho '}=$237 MeV, and $\Gamma_{\rho ''}=$235 MeV and respective strength $N_\rho=1$, $N_{\rho '}=1.802\times 10^{-3}$, and $N_{\rho ''}=5.93\times 10^{-3}$~\cite{Asakawa:1993kb}.

Expressing the relative velocity as $v_{\rm rel}=|{\bf k}_1/E_1-{\bf k}_2/E_2|$, Eq. (\ref{rate1}) can be written in Lorentz-invariant form as
\begin{eqnarray}
\frac{dN}{d^4x}&=&\frac{1}{(2\pi)^6}\int \frac{d^3{\bf k}_1}{E_1}\frac{d^3{\bf k}_1}{E_2}f(k_1)f(k_2)\nonumber\\
&&~~~~~\times\frac{M^2}{2}\sqrt{1-\frac{4m_i^2}{M^2}}\sigma(M^2),
\label{rate2}
\end{eqnarray}
where $m_i$ is the mass of colliding particles.
Changing variables into $P_\mu=k_{1\mu}+k_{2\mu}$ and $k_\mu=(k_{1\mu}-k_{2\mu})/2$, we can rewrite the Lorentz-invariant phase space as
\begin{eqnarray}
\frac{d^3{\bf k}_1}{E_1}\frac{d^3{\bf k}_2}{E_2}=\frac{d^3{\bf P} d^3{\bf k}}{E_1 E_2}=\frac{d^3{\bf P}}{E}\frac{Ed^3{\bf k}}{E_1 E_2},
\end{eqnarray}
with $E=E_1+E_2$. Because $d^3{\bf P}/E$ is Lorentz invariant, $E d^3{\bf k}/(E_1 E_2)$ also should be Lorentz invariant. Boosting to the center-of-mass frame of the two particles, we obtain
\begin{eqnarray}
\frac{d^3{\bf P}}{E}\frac{Ed^3{\bf k}}{E_1 E_2}=\frac{d^3{\bf P}}{E}\frac{E'd^3{\bf k}^\prime}{E_1' E_2'}=\frac{d^3{\bf P}}{E}\frac{4d^3{\bf k}^\prime}{M},
\label{variables}
\end{eqnarray}
where ${\bf k}^\prime$ is the momentum, and
$E_1'$, $E_2'$ and $E'$ are energies of particles 1 and 2, and their total energy
in their center-of-mass frame, respectively.

Using Eq.(\ref{vdis}), the product of particle distribution functions becomes
\begin{eqnarray}
&&f(k_1)f(k_2)=f_0(k_1)f_0(k_2)+f_0(k_1)\delta f(k_2)\nonumber\\
&&~~~~~~~~~~~~~~~+f_0(k_2)\delta f(k_1)+\delta f(k_1)\delta f(k_2)\nonumber\\
\nonumber\\
&&=g_1g_2e^{-P\cdot u/T}\bigg[1+\frac{k_2^\mu k_2^\nu\pi_{\mu\nu}}{2T^2(e+p)}+\frac{k_1^\mu k_1^\nu\pi_{\mu\nu}}{2T^2(e+p)}\nonumber\\
&&~~~~~~~~~~~~~~~~~~~~~~~+\frac{k_1^\mu k_1^\nu k_2^\sigma k_2^\lambda\pi_{\mu\nu}\pi_{\sigma\lambda}}{4T^4(e+p)^2}\bigg],
\label{fdf}
\end{eqnarray}
where $g_1$ and $g_2$ are degeneracies of particle 1 and 2, respectively. Substituting Eqs. (\ref{variables}) and (\ref{fdf}) into Eq. (\ref{rate2}), we obtain
\begin{eqnarray}
&&\frac{dN}{d^4x}=\frac{2 g_1 g_2}{(2\pi)^6}\int\frac{d^3{\bf P}}{E} \int d^3{\bf k}^\prime e^{-P\cdot u/T}\bigg[1+\frac{k^{\mu'} k^{\nu'}\pi_{\mu\nu}^{'}}{T^2(e+p)}\nonumber\\
&&~~+\frac{k^{\mu'} k^{\nu'} k^{\sigma''} k^{\lambda''} \pi_{\mu\nu}'\pi_{\sigma\lambda}'}{4T^4(e+p)^2}\bigg]M\sqrt{1-\frac{4m_i^2}{M^2}}\sigma(M^2),
\label{rate3}
\end{eqnarray}
where $k^{\mu'}=(k_0',\vec{k}')$, $k^{\mu''}=(k_0',-\vec{k}')$ and $\pi_{\mu\nu}^{'}$ are momenta of particles 1 and 2, and the shear tensor in their center of mass frame, respectively. By using $d^4x=\tau d\tau r dr d\eta d\phi$ and $d^3{\bf P}/E=\pi dy dP_T^2$ and integrating over the solid angle of ${\bf k}^\prime$ in the right hand side of Eq. (\ref{rate3}), the differential yield of dileptons in heavy-ion collisions is then given by

\begin{eqnarray}
\frac{dN}{dydM^2dP_T^2}=\frac{g_1 g_2}{4(2\pi)^3}XM^2\sigma(M^2)~~~~~~~~~~~~~~~~~~~\nonumber\\
\times\int d\tau \tau ~dr r \bigg[I_0(\alpha)K_0(\beta)~~~~~~~~~~~~~~~~~~~~~~\nonumber\\
+\frac{M^2(1+X/3)}{4T^2(e+p)}\sum_{i,j=0,1,2}^{(i+j=0,2)}A_{ij}I_i(\alpha)K_j(\beta)\nonumber\\
+\frac{M^4}{64T^4(e+p)^2}\sum_{i,j=0,1,2,3,4}^{(i+j=0,2,4)}B_{ij}I_i(\alpha)K_j(\beta)\bigg].
\label{rate4}
\end{eqnarray}
In the above, $X=1-4m_i^2/M^2$, and $I_i(\alpha)$ and $K_j(\beta)$ are modified Bessel functions with $\alpha=P_T \sinh\rho/T$ and $\beta=M_T \cosh\rho/T$, where $\rho=\tanh^{-1}v_r$. The coefficients $A_{ij}$, $B_{ij}$ and details on the derivation of above expression are given in the Appendix.

\begin{figure}[h]
\centerline{
\includegraphics[width=9 cm]{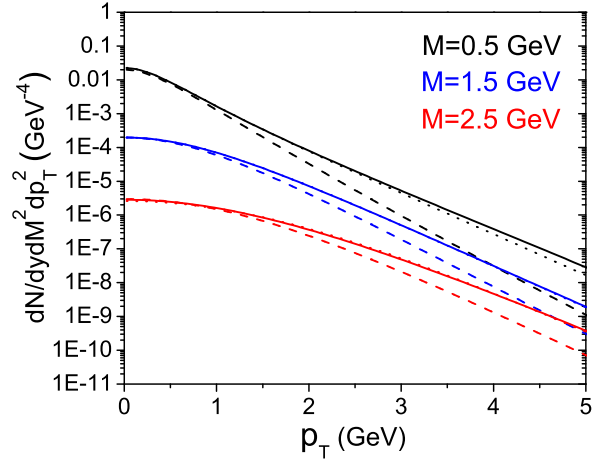}}
\caption{(color online) Transverse momentum spectra of dileptons of various invariant masses in viscous (solid lines) and ideal (dashed lines) hydrodynamics. Dotted lines correspond to results based on only the first-order correction from the modified particle transverse momentum distributions in the viscous hydrodynamics.}
\label{dilepton-pt}
\end{figure}

In Fig. \ref{dilepton-pt}, we show by solid lines the transverse momentum ($p_T$) spectrum of dileptons from the viscous hydrodynamics for various dilepton invariant masses. Also shown by dotted lines are those including only the first-order correction from the modified particle transverse momentum distributions, i.e., the second term in Eq.(\ref{rate4}), and they are seen to be very close to those including also the second-order correction given by the third term in Eq.(\ref{rate4}).
Compared with those from the ideal hydrodynamics given by the first term of Eq.(\ref{rate4}) and shown by dashed lines, the dilepton spectra in the viscous hydrodynamics are enhanced at high $p_T$ as in Ref.~\cite{Dusling:2008xj}. The viscous effect on the dilepton transverse momentum spectra is thus similar to that on the particle transverse momentum spectra as a result of enhanced density of quarks in QGP or pions in HG at high $p_T$. Since dileptons of small invariant masses are mainly produced from pion-pion annihilations in HG, in which the viscosity is large, the viscous effect is thus particularly large for dileptons of small invariant masses. Too large a viscous correction makes, however, the results from the viscous hydrodynamics unreliable. In fact, negative values can appear for the dilepton yield from pion-pion annihilation in integrating Eq. (\ref{rate4}) over the radial variable $r$ if the dilepton invariant mass is large. In this case, we have set the negative value to zero. This has, however, a negligible effect on the final result as the contribution of pion-pion annihilations in HG to dileptons of large invariant masses is insignificant compared to that from quark-antiquark annihilations in QGP.

\begin{figure}[h]
\centerline{
\includegraphics[width=9 cm]{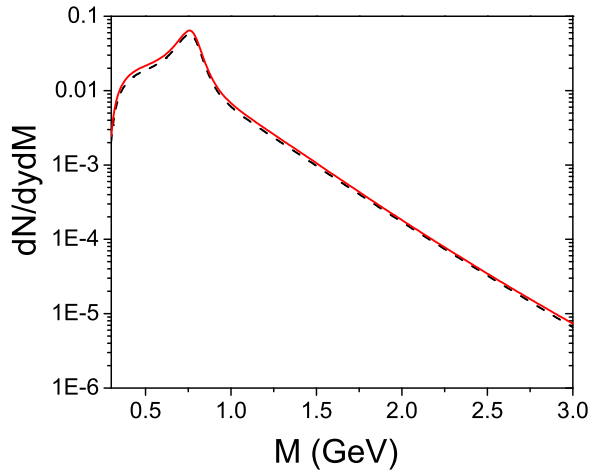}}
\caption{Dilepton spectra from viscous (solid line) and ideal hydrodynamics (dashed line).}
\label{spectrum}
\end{figure}

Figure \ref{spectrum} shows the dilepton invariant mass spectra from the viscous (solid line) and the ideal (dashed line) hydrodynamics. The two are seen to be almost identical at all invariant masses as the dilepton transverse momentum spectra at a fixed invariant mass are similar at low transverse momenta, which dominate the dilepton yield, in both the viscous and the ideal hydrodynamics.

\begin{figure}[h]
\centerline{
\includegraphics[width=9 cm]{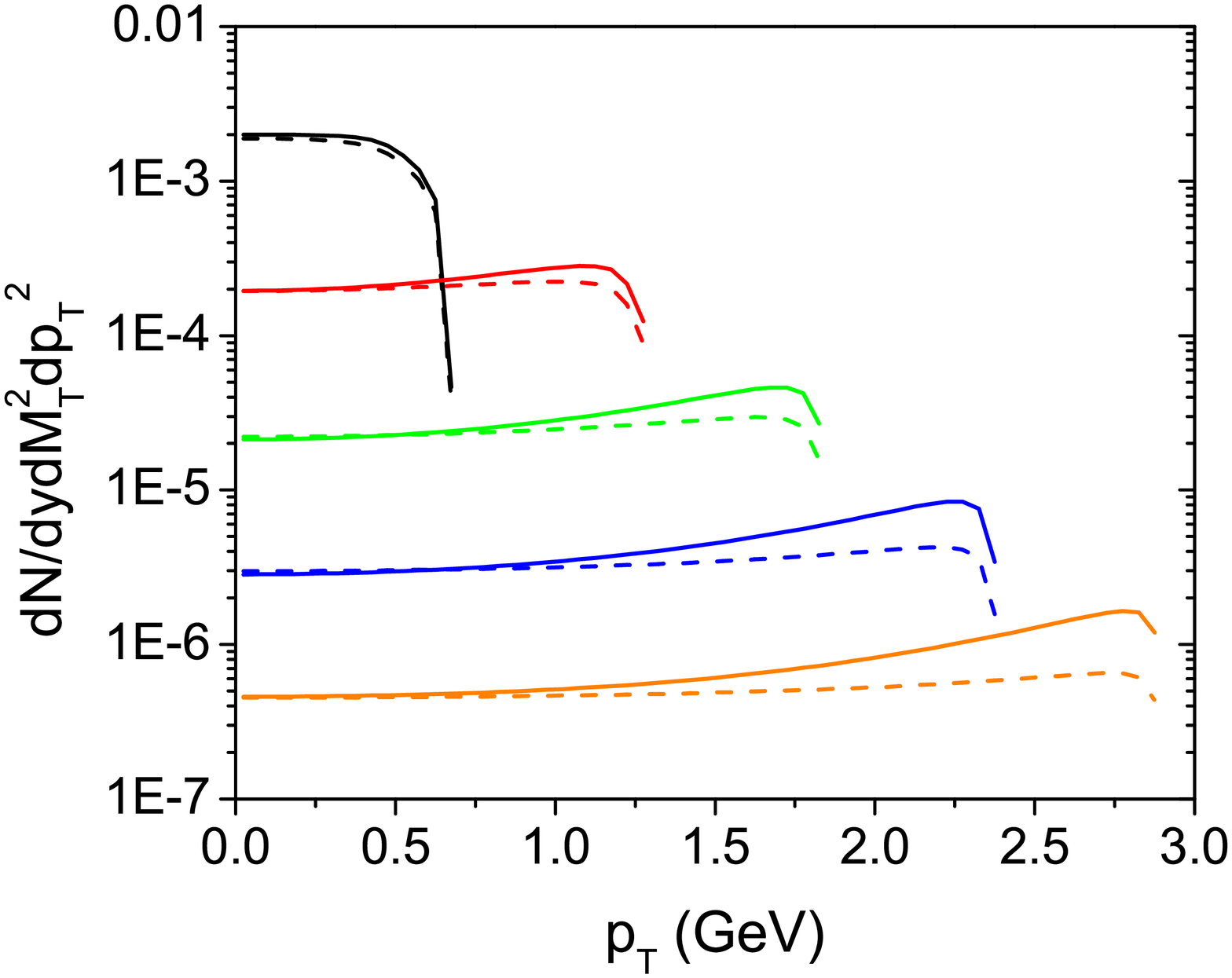}}
\caption{Transverse momentum spectra of dileptons from QGP with
transverse mass $M_T=$1, 1.5, 2, 2.5 and 3 GeV from top to bottom. Dashed and solid lines are, respectively, from ideal and viscous hydrodynamics.}
\label{mT-scaling}
\end{figure}

In Ref.~\cite{Asakawa:1993kb}, it has been shown in the ideal hydrodynamics that the differential yield of thermal dileptons with a fixed value of transverse mass $M_T$ is essentially independent of their transverse momenta if they are produced from QGP \cite{Asakawa:1993kb}, although this so-called $M_T$-scaling is violated for dileptons produced from HG. In Fig. \ref{mT-scaling}, we show by dashed lines the transverse momentum spectra of dileptons with $M_T=$1, 1.5, 2, 2.5 and 3 GeV from QGP in the ideal hydrodynamics. It is seen that the $M_T$ scaling still holds approximately, although we have used massive quarks and gluons in our study while massless ones were used in Ref.~\cite{Asakawa:1993kb}. The results from the viscous hydrodynamics are shown by solid lines and it shows that including viscosity leads to a violation of the $M_T$ scaling of dileptons from QGP at high $P_T$.

\section{Discussions and summary}\label{summary}

We have derived a set of schematic equations from the causal viscous hydrodynamics of Israel-Stewart
for central relativistic heavy-ion collisions by assuming that not only the energy density, pressure and entropy density are uniform in the produced fire-cylinder but also the azimuthal and space-time rapidity components of the shear tensor. Solving these equations using the massive quasi-particle model for the equation of state of QGP and the resonance gas model for that of HG, we have found that the shear viscosity slightly delays the cooling of produced hot matter and enhances somewhat its transverse expansion. It also increases significantly the particle distributions at high $p_T$, compared with those in the ideal hydrodynamics. Using this model, we have investigated thermal dilepton production in relativistic heavy-ion collisions by including contributions from the dominant quark-antiquark and pion-pion annihilations. Because of the viscous effect, the dilepton $p_T$ spectrum is enhanced at high $p_T$, which is similar to those found in Ref.~\cite{Dusling:2008xj} based on the non-causal Navier-Stokes viscous hydrodynamics and the first-order viscous correction from the modified particle transverse momentum distributions. For the invariant mass spectrum of dileptons, it is found to differ very little from that in the ideal hydrodynamics. We have also studied the effect of viscosity on the $M_T$ scaling of thermal dileptons from QGP, i.e., the yield is independent of $p_T$ for fixed dilepton transverse mass $M_T$, that has been previously predicted in the ideal hydrodynamics with massless quarks and gluons, and it is found that the $M_T$ scaling still holds in the ideal hydrodynamics even if QGP is composed of massive quarks and gluons as a result of their strong couplings. The $M_T$ scaling of dileptons is, however, broken in the viscous hydrodynamics due to the enhancement in the number density of quarks and antiquarks at high $p_T$ by the viscous effect.

\section*{Acknowledgements}
This work was supported in part by the U.S. National Science Foundation under Grant No. PHY-0758115 and the Welch Foundation under Grant No. A-1358.


\hfil\break
\appendix
\centerline{\bf \large Appendix}
\bigskip

In this Appendix, we give the details on the derivation from Eq. (\ref{rate3}) to (\ref{rate4}). Keeping the non-vanishing terms in the integration with respect to ${\bf k}^\prime$ in the right hand side of Eq. (\ref{rate3}) leads to
\begin{widetext}
\begin{eqnarray}
\frac{dN}{d^4x}&=&\frac{2 g_1 g_2}{(2\pi)^6}\int\frac{d^3 {\bf P}}{E} \int d^3{\bf k}^\prime M\sqrt{1-\frac{4m_i^2}{M^2}}\sigma(M^2) e^{-P\cdot u/T}\bigg[1+\frac{{k'}_0^2 \pi_{00}'+{k'}_1^2 \pi_{11}'+{k'}_2^2 \pi_{22}'+{k'}_3^2 \pi_{33}'}{T^2(e+p)}\nonumber\\
&&+\frac{1}{4T^4(e+p)^2}\bigg\{{k'}_0^4 {\pi_{00}'}^2+{k'}_1^4 {\pi_{11}'}^2+{k'}_2^4 {\pi_{22}'}^2+{k'}_3^4 {\pi_{33}'}^2\nonumber\\
&&~~~~~~+4(-{k'}_0^2 {k'}_1^2{\pi_{01}'}^2-{k'}_0^2 {k'}_2^2{\pi_{02}'}^2-{k'}_0^2 {k'}_3^2{\pi_{03}'}^2+{k'}_1^2 {k'}_2^2{\pi_{12}'}^2+{k'}_2^2 {k'}_3^2{\pi_{23}'}^2+{k'}_3^2 {k'}_1^2{\pi_{31}'}^2)\nonumber\\
&&+2({k'}_0^2 {k'}_1^2 \pi_{00}'\pi_{11}'+{k'}_0^2 {k'}_2^2 \pi_{00}'\pi_{22}'+{k'}_0^2 {k'}_3^2 \pi_{00}'\pi_{33}'+{k'}_1^2 {k'}_2^2 \pi_{11}'\pi_{22}'+{k'}_2^2 {k'}_3^2 \pi_{22}'\pi_{33}'+{k'}_3^2 {k'}_1^2 \pi_{33}'\pi_{11}')\bigg\}\bigg].
\end{eqnarray}
Splitting the integral $d^3 {\bf k}^\prime$ into the radial and angular parts according to
\begin{eqnarray}
d^3 {\bf k}^\prime=k'^2 dk'd\Omega=\frac{M}{16}\sqrt{1-\frac{4m_i^2}{M^2}}dM^2 d\Omega,
\end{eqnarray}
and integrating with respect to the solid angle $d\Omega$, we obtain
\begin{eqnarray}
\frac{dN}{d^4x}&=&\frac{g_1 g_2}{4(2\pi)^5}\int\frac{d^3 {\bf P}}{E} \int dM^2 X M^2\sigma(M^2) e^{-P\cdot u/T}\bigg[1+\bigg(\frac{M^2}{4}\bigg)\frac{\pi_{00}'+(X/3)(\pi_{11}'+\pi_{22}'+\pi_{33}')}{T^2(e+p)}\nonumber\\
&&+\frac{1}{4T^4(e+p)^2}\bigg(\frac{M^2}{4}\bigg)^2\bigg\{{\pi_{00}'}^2+\frac{2X}{3}\bigg(\pi_{00}'\pi_{11}'+\pi_{00}'\pi_{22}'+\pi_{00}'\pi_{33}'-2({\pi_{01}'}^2+{\pi_{02}'}^2+{\pi_{03}'}^2)\bigg)\nonumber\\
&&~~~~~~~~~~+\frac{X^2}{15}\bigg(3({\pi_{11}'}^2+{\pi_{22}'}^2+{\pi_{33}'}^2) +4({\pi_{12}'}^2+{\pi_{23}'}^2+{\pi_{31}'}^2)+2(\pi_{11}'\pi_{22}'+\pi_{22}'\pi_{33}'+\pi_{33}'\pi_{11}')\bigg)\bigg\}\bigg],
\label{pre}
\end{eqnarray}
where $X=1-4m_i^2/M^2$. Using the traceless property of the shear tensor $\pi^\mu_\mu=0$, Eq. (\ref{pre}) can be rewritten as

\begin{eqnarray}
\frac{dN}{dydM^2dP_T^2}&=&\frac{g_1 g_2}{8(2\pi)^4}XM^2\sigma(M^2)\int d\tau \tau ~dr r \int d\eta ~d\phi ~e^{-P\cdot u/T}\bigg[1+\frac{M^2(1+X/3)}{4T^2(e+p)}\pi_{00}'\nonumber\\
&&~~~~~~~~~~+\frac{M^4}{64T^4(e+p)^2}\bigg\{{\pi_{00}'}^2+\frac{2X}{3}\bigg({\pi_{00}'}^2-2({\pi_{01}'}^2+{\pi_{02}'}^2+{\pi_{03}'}^2)\bigg)\nonumber\\
&&~~~~~~~~~~~~~~~~~~~~~~~~~~~~~~~+\frac{X^2}{15}\bigg({\pi_{00}'}^2+2({\pi_{11}'}^2+{\pi_{22}'}^2+{\pi_{33}'}^2) +4({\pi_{12}'}^2+{\pi_{23}'}^2+{\pi_{31}'}^2)\bigg)\bigg\}\bigg].
\label{final}
\end{eqnarray}
\end{widetext}

The above integrals can be evaluated by transforming the shear tensor in the center of mass frame $\pi_{\mu\nu}'$ to the shear tensor in the fire-cylinder frame $\pi_{\mu\nu}$ through the Lorentz transformation, i.e.,
\begin{eqnarray}
\pi_{\mu\nu}'=\bigg(a_{\mu 0}\frac{\partial t}{\partial \tau}+a_{\mu 3}\frac{\partial z}{\partial \tau}\bigg)\bigg(a_{\nu 0}\frac{\partial t}{\partial \tau}+a_{\nu 3}\frac{\partial z}{\partial \tau}\bigg)\pi_{\tau\tau}~~~~\nonumber\\
+\bigg(a_{\mu 0}\frac{\partial t}{\partial \eta}+a_{\mu 3}\frac{\partial z}{\partial \eta}\bigg)\bigg(a_{\nu 0}\frac{\partial t}{\partial \eta}+a_{\nu 3}\frac{\partial z}{\partial \eta}\bigg)\pi_{\eta\eta}~~~~\nonumber\\
+\bigg(a_{\mu 1}\frac{\partial x}{\partial r}+a_{\mu 2}\frac{\partial y}{\partial r}\bigg)\bigg(a_{\nu 1}\frac{\partial x}{\partial r}+a_{\nu 2}\frac{\partial y}{\partial r}\bigg)\pi_{rr}~~~~\nonumber\\
+\bigg(a_{\mu 1}\frac{\partial x}{\partial \phi}+a_{\mu 2}\frac{\partial y}{\partial \phi}\bigg)\bigg(a_{\nu 1}\frac{\partial x}{\partial \phi}+a_{\nu 2}\frac{\partial y}{\partial \phi}\bigg)\pi_{\phi\phi}~~~\nonumber\\
+\bigg[\bigg(a_{\mu 0}\frac{\partial t}{\partial \tau}+a_{\mu 3}\frac{\partial z}{\partial \tau}\bigg)\bigg(a_{\nu 1}\frac{\partial x}{\partial r}+a_{\nu 2}\frac{\partial y}{\partial r}\bigg)~~~~~~~~\nonumber\\
+\bigg(a_{\nu 0}\frac{\partial t}{\partial \tau}+a_{\nu 3}\frac{\partial z}{\partial \tau}\bigg)\bigg(a_{\mu 1}\frac{\partial x}{\partial r}+a_{\mu 2}\frac{\partial y}{\partial r}\bigg)\bigg]\pi_{\tau r}~~~\nonumber\\
\nonumber\\
=(a_{\mu 0}\cosh\eta +a_{\mu 3}\sinh\eta)(a_{\nu 0}\cosh\eta+a_{\nu 3}\sinh\eta)\pi_{\tau\tau}\nonumber\\
+(a_{\mu 0}\sinh\eta +a_{\mu 3}\cosh\eta)(a_{\nu 0}\sinh\eta +a_{\nu 3}\cosh\eta)\tau^2\pi_{\eta\eta}\nonumber\\
+(a_{\mu 1}\cos\phi +a_{\mu 2}\sin\phi)(a_{\nu 1}\cos\phi +a_{\nu 2}\sin\phi)\pi_{rr}\nonumber\\
+(a_{\mu 1}\sin\phi -a_{\mu 2}\cos\phi)(a_{\nu 1}\sin\phi-a_{\nu 2}\cos\phi)r^2\pi_{\phi\phi}\nonumber\\
+[(a_{\mu 0}\cosh\eta+a_{\mu 3}\sinh\eta)(a_{\nu 1}\cos\phi+a_{\nu 2}\sin\phi)\nonumber\\
+(a_{\nu 0}\cosh\eta+a_{\nu 3}\sinh\eta)(a_{\mu 1}\cos\phi +a_{\mu 2}\sin\phi)]\pi_{\tau r},\nonumber\\
\label{shear9}
\end{eqnarray}
where $a_{\mu\nu}$ are components of the Lorentz transformation matrix,
\begin{eqnarray}
a_{00}&=&(M_T/M)\cosh y\nonumber\\
a_{01}&=&a_{10}=-(P_T/M)\cos\phi_p\nonumber\\
a_{02}&=&a_{20}=-(P_T/M)\sin\phi_p\nonumber\\
a_{03}&=&a_{30}=-(M_T/M)\sinh y\nonumber\\
a_{11}&=&1+\frac{(P_T/M)^2\cos^2\phi_p}{(M_T/M)\cosh y+1}\nonumber\\
a_{22}&=&1+\frac{(P_T/M)^2\sin^2\phi_p}{(M_T/M)\cosh y+1}\nonumber\\
a_{33}&=&1+\frac{(M_T/M)^2\sinh^2 y}{(M_T/M)\cosh y+1}\nonumber\\
a_{12}&=&a_{21}=\frac{(P_T/M)^2\cos\phi_p\sin\phi_p}{(M_T/M)\cosh y+1}\nonumber\\
a_{23}&=&a_{32}=\frac{(P_T/M)\cos\phi_p(M_T/M)\sinh y}{(M_T/M)\cosh y+1}\nonumber\\
a_{31}&=&a_{13}=\frac{(P_T/M)\sin\phi_p(M_T/M)\sinh y}{(M_T/M)\cosh y+1},
\end{eqnarray}
with $\eta$, $y$, $\phi$ and $\phi_p$ being the space-time rapidity, energy-momentum rapidity, azimuthal angle in configuration space and in momentum space, respectively. It is straightforward to show that the square brackets in Eq. (\ref{final}) is a function of $y-\eta$ and $\phi-\phi_p$ and leads to modified Bessel functions after integration over $\eta$ and $\phi_p$ as given by Eq.(\ref{rate4}), where the coefficients $A_{ij}$ and $B_{ij}$ are
\begin{eqnarray}
A_{00}&=&\frac{x^2}{2}(\pi_{\tau\tau}-\pi^\eta_\eta)+\frac{x^2-1}{2}(\pi_{rr}+\pi^\phi_\phi),\\
A_{02}&=&\frac{x^2}{2}(\pi_{\tau\tau}+\pi^\eta_\eta),\\
A_{11}&=&-2x\sqrt{x^2-1}~\pi_{\tau r},\\
A_{20}&=&\frac{x^2-1}{2}(\pi_{rr}-\pi^\phi_\phi),
\end{eqnarray}
\begin{widetext}
\begin{eqnarray}
B_{00}&=&\frac{1}{8}\bigg[3x^4-x^2(3x^2-8)\frac{2X}{3}+(9x^4-16x^2+16)\frac{X^2}{15}\bigg](\pi_{\tau\tau}^2+{\pi^\eta_\eta}^2)\nonumber\\
&&+\frac{1}{8}\bigg[3(x^2-1)^2-(x^2-1)(3x^2+5)\frac{2X}{3}+(9x^4-2x^2+9)\frac{X^2}{15}\bigg](\pi_{rr}^2+{\pi^\phi_\phi}^2)\nonumber\\
&&+\bigg[x^2(x^2-1)-(x^4-x^2+1)\frac{2X}{3}+(3x^4-3x^2-2)\frac{X^2}{15}\bigg]\pi_{\tau r}^2\nonumber\\
&&+\frac{1}{4}\bigg(1-\frac{2X}{3}+\frac{X^2}{5}\bigg)\bigg[-x^4\pi_{\tau\tau}\pi^\eta_\eta+(x^2-1)^2\pi_{rr}\pi^\phi_\phi+2x^2(x^2-1)(\pi_{\tau\tau}-\pi^\eta_\eta)(\pi_{rr}+\pi^\phi_\phi)\bigg],\\
B_{02}&=&\frac{x^2}{2}\bigg[x^2-(x^2-2)\frac{2X}{3}+(3x^2-4)\frac{X^2}{15}\bigg](\pi_{\tau\tau}^2-{\pi^\eta_\eta}^2)\nonumber\\
&&+x^2\bigg[x^2-1-x^2\frac{2X}{3}+(3x^2-1)\frac{X^2}{15}\bigg]\pi_{\tau r}^2\nonumber\\
&&+\frac{1}{2}\bigg(1-\frac{2X}{3}+\frac{X^2}{5}\bigg)x^2(x^2-1)(\pi_{\tau\tau}+\pi^\eta_\eta)(\pi_{rr}+\pi^\phi_\phi),\\
B_{11}&=&x\sqrt{x^2-1}\bigg[-3x^2+(3x^2-4)\frac{2X}{3}-(9x^2-8)\frac{X^2}{15}\bigg]\pi_{\tau r}\pi_{\tau\tau}\nonumber\\
&&+x\sqrt{x^2-1}\bigg[-3(x^2-1)+(3x^2+1)\frac{2X}{3}-(9x^2-1)\frac{X^2}{15}\bigg]\pi_{\tau r}\pi_{rr}\nonumber\\
&&+\bigg(1-\frac{2X}{3}+\frac{X^2}{5}\bigg)x\sqrt{x^2-1}~[x^2\pi^\eta_\eta-(x^2-1)\pi^\phi_\phi]\pi_{\tau r},\\
B_{20}&=&\frac{x^2-1}{2}\bigg[x^2-1-(x^2+2)\frac{2X}{3}+(3x^2+1)\frac{X^2}{15}\bigg](\pi_{rr}^2-{\pi^\phi_\phi}^2)\nonumber\\
&&+(x^2-1)\bigg[x^2-(x^2-1)\frac{2X}{3}+(3x^2-2)\frac{X^2}{15}\bigg]\pi_{\tau r}^2\nonumber\\
&&+\frac{1}{2}\bigg(1-\frac{2X}{3}+\frac{X^2}{5}\bigg)x^2(x^2-1)(\pi_{\tau\tau}-\pi^\eta_\eta)(\pi_{rr}-\pi^\phi_\phi),\\
B_{04}&=&\frac{1}{8}\bigg(1-\frac{2X}{3}+\frac{X^2}{5}\bigg)x^4(\pi_{\tau\tau}+\pi^\eta_\eta)^2,\\
B_{13}&=&-\bigg(1-\frac{2X}{3}+\frac{X^2}{5}\bigg)x^3\sqrt{x^2-1}(\pi_{\tau\tau}+\pi^\eta_\eta)\pi_{\tau r},\\
B_{22}&=&\frac{1}{2}\bigg(1-\frac{2X}{3}+\frac{X^2}{5}\bigg)x^2(x^2-1)\bigg[2\pi_{\tau r}^2+(\pi_{\tau\tau}+\pi^\eta_\eta)(\pi_{rr}-\pi^\phi_\phi)\bigg],\\
B_{31}&=&-\bigg(1-\frac{2X}{3}+\frac{X^2}{5}\bigg)x(x^2-1)^{3/2}(\pi_{rr}-\pi^\phi_\phi)\pi_{\tau r},\\
B_{40}&=&\frac{1}{8}\bigg(1-\frac{2X}{3}+\frac{X^2}{5}\bigg)(x^2-1)^2(\pi_{rr}-\pi^\phi_\phi)^2
\end{eqnarray}
\end{widetext}
with $x=M_T/M$.


\end{document}